\documentclass[12pt,a4paper]{article}
 \usepackage[centertags]{amsmath}
 \usepackage{amssymb}
 \usepackage{amstext}

\newcommand\chemp{\lambda}
\newcommand\rchem{\mu}
\newcommand\npart{\mathcal{N}}
\newcommand\msubs{\mathcal{M}}
\newcommand\corrf{C}
\newcommand{\mapping}{\mathop{\mathrm{arccot}}\nolimits}
\newcommand{\gcrit}{\mathop{g_{\mathrm{cr}}}\nolimits}
\newcommand{\GCRIT}{\mathop{G_{\mathrm{cr}}}\nolimits}
\newcommand\GCRITp{\mathop{G_{\mathrm{cr}}^{\prime}}\nolimits}
\newcommand\GCRITpp{\mathop{G_{\mathrm{cr}}^{\prime\prime}}\nolimits}
\newcommand\degen{S}
\newcommand\dispers{\mathcal{V}}
\newcounter{remm}
\newcommand{\remar}{\par\addtocounter{remm}{1}{\em Remark \arabic{remm}.\/} }
\begin{document}
 \begin{center}
 {\large
 {\bf
 Formation of superconducting pair correlations \\
 in spherical even-even nuclei

 }

 \medskip
 V.\,A.\ Kuz'min${}^{a)}$ and T.\,V.\ Tetereva${}^{b)}$

 }

 \medskip
 {\it
  ${}^{a)}$ Bogoliubov Laboratory of Theoretical Physics,
       Joint Institute \\ for Nuclear Research,
        Dubna, Moscow region, 141980, Russia  \\
   ${}^{b)}$ Skobeltsyn Institute of Nuclear Physics,
   Lomonosov  Moscow \\ State University, Moscow, 119992, Russia

 }

 \end{center}

 \abstract{ \noindent
  The appearance of like nucleon pair correlations in
  the ground state of spherical even-even nuclei is considered
  within a special Bogoliubov transformation.
  It is confirmed that in closed subshell nuclei superconducting
  pair correlations start to form if the coupling constant $G$
  exceeds a certain threshold value.
  Rough upper and lower estimates are obtained
  for the threshold value.
  It is shown, that superconducting correlations exist in
  open subshell nuclei at any positive $G$.
  In this case, nucleon pairs are distributed over all
  subshells participating in the pairing interaction.
 }

 \section*{Introduction}
 It was stressed in one of the very first papers \cite{Bel1959} on
 the application of the idea of superconducting correlations to
 the atomic nuclei spectroscopy
 that a pair correlations in nuclei can appear only
 if the coupling constant exceeds a certain value.
 It was also mentioned that the existence of such a threshold value
 distinguishes atomic nuclei from the infinite systems considered
 in the theory of superconductivity.
 This remark was later repeated several times \cite{VGS1971,BCS50}.

 The necessity of a threshold was revealed by the following
 reasoning \cite{Bel1959,VGS1971}.
 In the simplest case when all the matrix elements
 of the two-body pairing interactions are replaced by
 the same positive constant $G$, the equations for the
 correlation function $\corrf$ and the chemical potential $\chemp$
 can be written as
 \begin{gather}
   \label{eq:gap}
   \frac{G}{2} \sum_s \frac{1}{%
     \sqrt{ (E_s - \chemp)^2 + \corrf^2}} = 1 ; \\
  \label{eq:chem}
  2 \sum_s v_s^2 = \npart, \qquad
  v_s^2 = \frac{1}{2} \left\lbrack 1 -
   \frac{E_s - \chemp}{%
   \sqrt{ (E_s - \chemp)^2 + \corrf^2} } \right\rbrack .
 \end{gather}
 Here $\npart$ is the number of particles in the system.
 The summation is carried out over pairs of doubly-degenerate
 single particle states in the nuclear mean field.
 These states are related to each other by the operation
 of time reversal \cite{VGS1971}.
 The inequality
 \begin{equation}
 \label{eq:stb}
  \frac{G}{2} \sum_s \frac{1}{%
     \left \vert E_s - \chemp \right\vert } > 1 ,
 \end{equation}
 follows from Eq.\,(\ref{eq:gap}) for non-zero
 $\corrf$.
 Inequality (\ref{eq:stb}) is often considered
 as a relation that
 determines the minimal value of $G$, starting
 from which superconducting pair correlations can
 exist in the system.
 The reasoning contains a tacit assumption
 that $\chemp$ can not approach one of the $E_s$'s
 close enough to increase considerably the sum
 $ \displaystyle \sum_s 1/\left \vert E_s - \chemp \right\vert $
 as $G$ decreases.
 The correctness of the assumption is not evident because both
 $\chemp$ and $\corrf$ are calculated by solving
 the nonlinear system of Eqs.\ (\ref{eq:gap}) and (\ref{eq:chem}).
 Therefore, inequality (\ref{eq:stb}) is a useful hint
 rather than a proof.

 In this paper, we study the conditions of the formation of
 superconducting pair correlations between like nucleons
 in the ground state of even-even spherical nuclei.
 Since
 Eqs.~(\ref{eq:gap}) and (\ref{eq:chem}) for many-level systems
 can only be solved numerically, we  first consider the
 solution of equations with one and two subshells and
 then use the obtained results in the discussion
 of the many-subshell problem.

 The paper contains the introduction, three sections and
 the conclusion.
 The equations for calculating normal and abnormal
 single-particle densities are derived in the first section.
 The contribution of pairing interaction to the single-particle
 energy is taken into account explicitly.
 The appearance of pair correlations in single subshell and
 two-subshell systems is discussed in the second
 section.
 The threshold value of $G$ is calculated for the system
 with closed subshell, too.
 Many-subshell nuclei are discussed in the third section.
 The main results of the paper are summed up in the conclusion.

 \section{Model Hamiltonian and main equations}
 The Hamiltonian of the neutrons (or protons) in the nucleus
 with spherical symmetry is chosen \cite{VGS1971} as
 \begin{multline}
 \label{eq:hamil}
  H = \sum_k \sum_{m_k = -j_k}^{j_k} E_k a^{\dag}_{k, m_k} a_{k,m_k} -
 \\
 - \frac{G}{4} \sum_{k,l} \sum_{m_k=-j_k}^{j_k}
  (-1)^{j_k - m_k}  a^{\dag}_{k, m_k} a^{\dag}_{k, -m_k}
 \sum_{m_l=-j_l}^{j_l}
  (-1)^{j_l - m_l}  a_{l, -m_l} a_{l, m_l} .
 \end{multline}
 Here the indices $k$ label single-particle states with energy
 $E_k$ and angular momentum $j_{k}$;
 $m_{k}$ is the third projection of angular momentum,
 $m_k = -j_k, -j_k +1,\dots j_k-1, j_k$;
 the operators $a^{\dag}_{k,m_k}$ and  $a_{k,m_k}$ are fermion operators
 of the creation and annihilation of a particle in the state $(k,m_k)$.
 We follow the tradition and call {\it the subshell\/} a set of  $2 j_k + 1$
 single-particle states with the same $E_k$ and $j_k$.
 We consider the attractive interaction, i.e.\ $G > 0$.

 The quasiparticle operators are introduced by the special Bogoliubov
 transformation with the real coefficients \cite{VGS1971}
 \begin{equation}
  \label{eq:bogol}
   a_{k,m_k} = u_k \alpha_{k, m_k}
   + (-1)^{j_k - m_k} v_k \alpha^{\dag}_{k, - m_k} .
 \end{equation}
 If the operators $\alpha_{k, m_k}$ and $\alpha^{\dag}_{k, m_k}$ are
 the fermion annihilation and creation operators, the transformation
 coefficients satisfy the conditions
 \begin{equation}
 \label{eq:canon}
   u^2_k + v^2_k = 1 .
 \end{equation}
 The quasiparticle vacuum $\vert \rangle$ is determined
 by the equations $ \alpha_{k, m_k} \vert \rangle = 0$
 for any $k$ and $m_k$.
 If the quasiparticle operators are introduced by
 Eq.\ (\ref{eq:bogol}), the angular momentum of $\vert \rangle$
 is zero and its parity is positive.
 This is easy to prove by applying the operators of the total
 angular momentum of the system of like nucleons and the parity
 operator on $\vert \rangle$.
 As a result, the product of neutron and proton vacuum states
 has zero angular momentum and positive parity,
 the values coincide with the angular momentum and parity of
 ground states of all even-even spherical nuclei.
 For this reason,
 we approximate the wave function of the ground state of a system
 containing an even number $\npart$ of
 identical nucleons (either neutrons or protons) by
 an appropriate quasiparticle vacuum $\vert \rangle$.

 The quasiparticle vacuum is connected with the state without
 nucleons \cite{VGS1971}:
 \begin{equation}
 \label{eq:vacu}
  \vert \rangle =
  \prod_{k, m_k > 0} \left ( u_k + (-1)^{j_k - m_k}
  v_k a^{\dag}_{k,m_k} a^{\dag}_{k,-m_k} \right )
  \vert 0 \rangle ,
 \end{equation}
 where $\vert 0 \rangle$ is the state without nucleons,
 $ a_{k, m_k} \vert 0 \rangle = 0$ for any $k$ and $m_k$.
 If for a certain $k$ the coefficients $u_k$ and $v_k$ differ
 simultaneously from zero ($ u_k v_k \neq 0 $), the function
 (\ref{eq:vacu})
 contains components with a different number
 of nucleon pairs.

 The coefficients of transformation (\ref{eq:bogol}) can be
 determined from the condition of minimum of the system energy
 $\langle \vert H \vert \rangle$.
 In searching for an energy extremum, one should satisfy the constraint
 that the average of the nucleon number operator over quasiparticle vacuum
 is equal to the number of particles in the system,
 \begin{equation*}
 \langle \vert N \vert \rangle =
 \langle \vert \sum_{k,m_k} a^{\dag}_{k,m_k} a_{k,m_k}
  \vert \rangle = \npart .
 \end{equation*}
 Therefore, the functional to be minimized is
 \begin{equation*}
  \langle \vert H \vert \rangle
  - \chemp \left \lbrack
  \langle \vert \sum_{k,m_k} a^{\dag}_{k,m_k} a_{k,m_k}
   \vert \rangle - \npart \right \rbrack.
 \end{equation*}
 Here $\chemp$ is the Lagrange multiplier usually called
 {\it the chemical potential.\/}
 One can see that the inclusion of a supplementary condition
 modifies the operator $H$ into
 \begin{equation*}
   H^{\prime} = H - \chemp  \sum_{k,m_k}
   a^{\dag}_{k,m_k} a_{k,m_k} .
 \end{equation*}
 The average of $H^{\prime}$ over the quasiparticle vacuum is equal to
 \begin{equation}
 \label{eq:average}
 \langle \vert H^{\prime} \vert \rangle =
  \sum_{k} ( 2 j_k + 1 ) ( E_k - \chemp ) v^2_{k}
  - \frac{G}{2} \sum_{k} (2 j_k + 1) v^4_k
  - \frac{G}{4} \left \lbrack \sum_{k}
 ( 2 j_k +1 ) u_k v_k \right \rbrack^{2} \!\!.
 \end{equation}
 The transformation coefficients $u_k$ and $v_k$ enter into
 the expression as the products $v^2_k$ and $u_k v_k$, only.
 We use the products as new variables
 \begin{equation}
 \label{eq:dens}
  w_k = v^2_k   \quad  \mathrm{and}
  \quad
  t_k = u_k v_k \,.
 \end{equation}
 Sometimes they are called the {\it normal density\/} and
 {\it abnormal density\/} because
 \begin{equation*}
  w_k = \langle \vert a^{\dag}_{k, m_k} a_{k, m_k} \vert \rangle
 \quad \mathrm{and} \quad
  t_k = (-1)^{j_k - m_k}
 \langle \vert a_{k, -m_k}  a_{k, m_k} \vert \rangle .
 \end{equation*}
 The presence of non-zero
 $\langle \vert a_{k, -m_k}  a_{k, m_k} \vert \rangle$
 is a sign of a superconducting state.
 It follows from Eq.\ (\ref{eq:canon}) that the densities
 obey to the inequalities
  \begin{equation*}
  0 \leqslant w_k \leqslant 1
  \quad \mathrm{and} \quad
  0 \leqslant t_k \leqslant \frac{1}{2}
  \end{equation*}
 and are connected by the relations
 \begin{equation}
 \label{eq:canondens}
  w^2_k + t^2_k = w_k \, .
 \end{equation}
 The matrix element (\ref{eq:average}) is expressed via
  $w_k$ and $t_k$ as
 \begin{equation}
 \label{eq:averagedens}
 \langle \vert H^{\prime} \vert \rangle =
  \sum_{k} ( 2 j_k + 1 ) ( E_k - \chemp ) w_{k}
  - \frac{G}{2} \sum_{k} (2 j_k + 1) w^2_k
  - \frac{G}{4} \left \lbrack \sum_{k}
 ( 2 j_k +1 ) t_k \right \rbrack^{2} .
 \end{equation}
 The supplementary condition for the particle number
 is
 \begin{equation}
 \label{eq:averagepart}
 \sum_k (2 j_k + 1 ) w_k = \npart .
 \end{equation}
 For convenience of further calculations, we write $w_k$ as
 a sum
 \begin{equation}
 \label{eq:circleone}
  w_k = \xi_k + \frac{1}{2},
  \qquad \left ( -1/2 \leqslant \xi_k \leqslant 1/2 \right ) .
 \end{equation}
 Now relations (\ref{eq:canondens}) are simplified
 \begin{equation}
 \label{eq:circletwo}
  \xi^2_k + t^2_k = \frac{1}{4}
 \end{equation}
 and can be satisfied by the trigonometric functions
 \begin{equation*}
  \xi_k = \frac{1}{2} \cos \varphi_k \,,
  \quad
  t_k = \frac{1}{2} \sin \varphi_k \,,
  \qquad
  \mathrm{with}
  \ \ 0 \leqslant \varphi_k \leqslant \pi .
 \end{equation*}
 Please note that $t_k > 0$ with $ 0 < \varphi_k < \pi $.
 The matrix element (\ref{eq:average}) depends on the unknown
 $\varphi_k$ only,
 \begin{multline}
 \langle \vert H^{\prime} \vert \rangle =
 \sum_k  \bigl ( j_k + \frac{1}{2} \bigr ) \bigl ( E_k - \chemp - \frac{G}{4} \bigr )
 + \sum_k \bigl ( j_k + \frac{1}{2} \bigr )
   \bigl ( E_k - \frac{G}{2} - \chemp \bigr ) \cos \varphi_k - \\
 - \frac{G}{4} \sum_k \bigl ( j_k + \frac{1}{2} \bigr ) \cos^2 \varphi_k
 - \frac{G}{4} \left \lbrack \sum_k \bigl ( j_k + \frac{1}{2} \bigr )
   \sin \varphi_k \right \rbrack ^2 .
 \end{multline}
 For any $\varphi_k$ from the interval
 $0 < \varphi_k < \pi$ the condition of extremum
 $ \langle \vert H^{\prime} \vert \rangle $ is
 \begin{equation}
  \label{eq:extremal}
  \frac{\partial \langle \vert H^{\prime} \vert \rangle}{%
  \partial \varphi_k } = 0 \, .
 \end{equation}
 When $\varphi_k$  approaches $ 0 $ or $ \pi $,
 the usual derivatives should be replaced by the one-sided ones.
 The simplest way to arrange the limiting processes
 is to map the entire number axis to
 the segment $ 0 \leqslant \varphi_k \leqslant \pi $
 by
 $ \varphi_k = \mapping x_k$ with $\infty > x_k  > -\infty $.
 Afterwards, Eq.\,(\ref{eq:extremal}) transforms into
 \begin{equation}
  \label{eq:extremalcomplex}
  \frac{\partial \langle \vert H^{\prime} \vert \rangle}{%
  \partial x_k } =
  \frac{\partial \langle \vert H^{\prime} \vert \rangle}{%
  \partial \varphi_k }
  \frac{d \varphi_k}{ d x_k} =
  \frac{\partial \langle \vert H^{\prime} \vert \rangle}{%
  \partial \varphi_k }
  \left ( - \frac{1}{1 + x^2_k } \right )
   = 0 \, .
 \end{equation}
 Three stationary points are possible for each $k$:
 \begin{gather}
 \label{eq:statone}
  x_k = \infty, \quad \varphi_k = 0, \quad \xi_k = \frac{1}{2}, \ w_k=1, \ t_k =0 ; \\
 \label{eq:stattwo}
  x_k = -\infty, \quad \varphi_k = \pi, \quad \xi_k = -\frac{1}{2}, \ w_k = 0, \  t_k =0 ; \\
 \label{eq:statthree}
   x_k \ \mathrm{finite} , \quad
  \frac{\partial \langle \vert H^{\prime} \vert \rangle}{%
  \partial \varphi_k } =0, \quad
  0 < \varphi_k < \pi ,  \quad t_k > 0 , \quad 0 < w_k < 1.
 \end{gather}
 Stationary points (\ref{eq:statone}) and (\ref{eq:stattwo})
 describe the states with normal density at subshell $k$ strictly
 equal to either $1$ or $0$ and abnormal density equal to zero
 (normal solutions).
 Equation (\ref{eq:statthree}) describes the states with $t_k > 0$
 and $0 < w_k < 1$ (superconducting solutions).
 Equation (\ref{eq:statthree}) can be written as
 \begin{equation*}
   \bigl ( E_k - \frac{G}{2} - \chemp \bigr ) \sin \varphi_k
  - \frac{G}{2} \cos \varphi_k \sin \varphi_k
  + \frac{G}{2} \left \lbrack \sum_l \bigl ( j_l + \frac{1}{2} \bigr )
    \sin \varphi_l \right \rbrack \cos \varphi_k = 0 ,
 \end{equation*}
 or
 \begin{equation}
 \label{eq:main}
  \bigl ( E_k - \frac{G}{2} - \chemp \bigr ) t_k
 - G \xi_k t_k + G D \xi_k = 0.
 \end{equation}
 Here the notation
 \begin{equation*}
  D = \sum_l \degen_l \, t_l \,,
 \quad
 \degen_l = j_l + \frac{1}{2} \,.
 \end{equation*}
 is used.
 For each subshell $l$ the quantities $\degen_l$ are equal
 to the number of different particle pairs that form
 the state (\ref{eq:vacu}).

 \remar
 Usually in Eq.\,(\ref{eq:average}) the terms proportional
 to $v^4_k$ are discarded (or absorbed into $E_k$ by the modification
 of single-particle energies \cite{VGS1971}).
 In this case, Eq.\,(\ref{eq:main}) is transformed into
 \begin{equation*}
  \bigl ( E_k - \chemp \bigr ) t_k = - G D \xi_k .
 \end{equation*}
 We square both parts of the equation, use Eq.\,(\ref{eq:circletwo})
 and obtain equations analogues to Eqs.\ (\ref{eq:gap})
 and (\ref{eq:chem}) with the correlation function $\corrf = G D$.

 \remar
 Extremum conditions (\ref{eq:statone}) and (\ref{eq:stattwo})
 at $\varphi_k = 0$ and  $\varphi_k = \pi$
 correspond to the normal solutions with $t_k = 0$.
 These solutions allow one to
 describe the system having $t_l > 0$ for a certain subshell and
  $t_m = 0$ for others subshells.
 If $t_{k^\prime} = 0 $ for a certain subshell $k^{\prime}$
 then from Eqs.\,(\ref{eq:main}) alone it would follow
 that $D = 0$ and the rest of the abnormal densities would be zero.

 \remar
 The contribution of components with a different particle number
 to $\vert \rangle$ is estimated by the
 particles number variance
 \begin{equation*}
  \dispers = \langle \vert N^2 \vert \rangle
   -  \langle \vert N \vert \rangle^2
  = 2 \sum_{k} ( 2 j_k + 1 ) \, t^2_k
  = 4 \sum_k \degen_k  t^2_k .
 \end{equation*}
 The particle number variance is determined by non-zero
 $ t_k $ only.

 \section{One and two subshells}
 Let us start with the simplest cases.

 \subsection{Single subshell}%
 \label{sec:single}
 If the system contains a single subshell
 with the energy $E_0$ and angular momentum $j_0$,
 Eqs.\ (\ref{eq:main}) and (\ref{eq:averagepart})
 are reduced to
 \begin{gather}
  \label{eq:oneleva}
  \bigl ( E_0 - \frac{G}{2} - \chemp \bigr ) t_0
 - G \xi_0 t_0 + G \degen_0 t_0 \xi_0 = 0, \\
  \label{eq:onelevb}
   2 \degen_0 \left ( \xi_0 + \frac{1}{2} \right ) = \npart.
 \end{gather}

 If $\npart = 2 \degen_0$ (all single-particle states are
 occupied or {\it the subshell is closed\/}), the particle number
 equation (\ref{eq:onelevb}) has the solution $ \xi_0 = 1/2 $,
 and therefore $t_0 = 0$.
 Only the normal solution is possible for any $G$.

 If $\npart = 2 \degen_0 - 2P_0$, here $P_0$ is the number
 of particle pairs removed from the closed subshell.
 The particle number equation (\ref{eq:onelevb}) has the solution
 \begin{equation*}
  \xi_0 = 1/2 - p_0 , \quad \mathrm{with} \ p_0 = P_0/\degen_0,
  \quad \mathrm{and} \quad w_0 = 1 - p_0 , \quad w_0 < 1 ,
 \end{equation*}
 and only superconducting solution exists.
 The abnormal density and correlation function are equal to
 \begin{gather*}
  t_0 = \sqrt{p_0 ( 1 - p_0 ) } , \\
  \corrf = G D = G \degen_0 t_0
  = G \degen_0 \sqrt{p_0(1-p_0)}
   = \frac{G}{2} \sqrt{\npart ( 2\degen_0 - \npart)} .
 \end{gather*}
 In the case of single open subshell, the correlation
 function depends linearly on $G$,
 because both $\xi_0$ and $t_0$ are independent
 of $G$.
 The particle number variance is also constant
  \begin{equation*}
  \dispers = 4 \degen_0 t^2_0 = 4 \degen_0 p_0 ( 1 - p_0 )
  = \frac{\npart ( 2 \degen_0 - \npart )}{\degen_0} .
 \end{equation*}
 The chemical potential is a linear function of $G$,
 \begin{equation*}
  \chemp = E_0 + G \left \lbrack
   \left ( \frac{1}{2} - p_0 \right )\degen_0
  - \left ( 1 -  p_0 \right ) \right \rbrack .
 \end{equation*}

 \subsection{Two subshells}
 For the system having two subshells, the set of
 equations is
 \begin{gather}
 \label{eq:twolevels}
 \begin{split}
  \bigl ( E_1 - \frac{G}{2} - \chemp \bigr ) t_1
 - G t_1 \xi_1 + G \left ( \degen_1 t_1 + \degen_2 t_2 \right ) \xi_1 & = 0, \\
  \bigl ( E_2 - \frac{G}{2} - \chemp \bigr ) t_2
 - G t_2 \xi_2 + G \left ( \degen_1 t_1 + \degen_2 t_2 \right ) \xi_2 & = 0,
 \end{split} \\
 \noalign{\smallskip}
 \label{eq:twosubparticles}
  2 \left ( \degen_1 \xi_1 + \degen_2 \xi_2 \right )
  = \npart - \degen_1 - \degen_2 .
 \end{gather}
 We label subshells by the indices "1" and "2" so that $E_2 > E_1$.
 If $E_2 = E_1$ we face, due to definition (\ref{eq:bogol}), the
 single subshell system with $\degen_0 = \degen_1 + \degen_2$.
 It is convenient to introduce the dimensionless chemical potential
 and coupling constant
  \begin{equation*}
   \rchem = \frac{\chemp - E_1 + G/2}{E_2-E_1}
   \quad
   \mathrm{and}
   \quad
    g = \frac{G}{E_2-E_1} .
  \end{equation*}
 And equations (\ref{eq:twolevels}) can be written as
  \begin{equation}
  \label{eq:dimtwo}
  \begin{aligned}
   - \rchem t_1
 - g t_1 \xi_1 + g \left ( \degen_1 t_1 + \degen_2 t_2 \right ) \xi_1 & = 0, \\
  \bigl ( 1 - \rchem \bigr ) t_2
 - g t_2 \xi_2 + g \left ( \degen_1 t_1 + \degen_2 t_2 \right ) \xi_2 & = 0.
  \end{aligned}
  \end{equation}

 \subsubsection{ $\degen_1 = \degen_2$ and $\npart = 2 \degen_1 $ }
  Consider the system with $\degen_1 = \degen_2$.
  The particle number equation (\ref{eq:twosubparticles})
  is simplified
  \begin{equation*}
  2 \degen_1 \left ( \xi_1 + \xi_2 \right ) = \npart - 2 \degen_1 .
  \end{equation*}
  When $\npart = 2 \degen_1 $, it gives
  $\xi_2 = -\xi_1$ and $t_2 = t_1$.
  Equations~(\ref{eq:dimtwo}) are transformed into
  \begin{equation}
   \label{eq:simplecombined}
    \begin{aligned}
   - \rchem t_1
  - g t_1 \xi_1 + 2 g \degen_1 t_1 \xi_1 & = 0, \\
  \bigl ( 1 - \rchem \bigr ) t_1
  + g t_1 \xi_1 - 2 g \degen_1 t_1 \xi_1 & = 0.
    \end{aligned}
  \end{equation}
 We sum them and get the equation
  \begin{equation*}
     (1 - 2 \rchem ) \, t_1 = 0,
  \end{equation*}
 having two solutions: $\rchem=1/2$ and $t_1 = 0$.

 The $t_1 = 0$ solution corresponds to the normal state
 with
  $\xi_1 = \pm 1/2$ and  $\xi_2 = \mp 1/2$.
 The energy minimum is reached at
 $\xi_1 = 1/2$
  ($w_1 = 1$) and $\xi_2 = -1/2$ ($w_2 = 0$).
 The solution exists for any positive $G$.
 The chemical potential $\rchem$ is arbitrary.

 Let us consider the $\rchem = 1/2$ solution with  $t_1 \neq 0$.
 The chemical potential is
  \begin{equation*}
   \chemp = \frac{1}{2} \left ( E_1 + E_2 - G \right ) .
  \end{equation*}
 It follows from the first of Eqs.\,(\ref{eq:simplecombined})
 for $t_1 \neq 0$ that
  \begin{equation*}
    \xi_1 = \frac{1}{2} \, \frac{1}{g \left ( 2 \degen_1 - 1 \right) } .
  \end{equation*}
 Due to Eq.\,(\ref{eq:circletwo}) the real non-zero $t_1$
 exists for $\vert \xi_1 \vert < 1/2 $;
 therefore, the superconducting pair correlations may appear if
  \begin{equation*}
    g > \frac{1}{2\degen_1 - 1} .
  \end{equation*}
 It is convenient to introduce special notation for the critical
  values of the interaction constant
 \begin{equation*}
    \gcrit = \frac{1}{2\degen_1 - 1}
     \quad \mathrm{and} \quad
     \GCRIT = \frac{E_2-E_1}{2\degen_1 - 1} .
 \end{equation*}
 The solutions are
 \begin{gather*}
  \xi_1 = \frac{1}{2} \frac{\gcrit}{g}
  = \frac{1}{2}\frac{\GCRIT}{G} , \\
  t_1 = \sqrt{\frac{1}{4} - \xi^2_1} =
  \frac{1}{2} \sqrt{\frac{(g - \gcrit)(g + \gcrit)}{g^2}} .
 \end{gather*}
 For small $g$, $ 0 < g - \gcrit \ll \gcrit $, we have
 (the approximate equality $g + \gcrit \approx 2 g $ is taken into account)
 $ \displaystyle t_1 \approx \sqrt{\frac{g - \gcrit}{2 g}} =
 \sqrt{\frac{G - \GCRIT}{2 G}} $,
 and $ \corrf = 2 G \degen_1 t_1 \approx \degen_1
  \sqrt{ 2 G (G - \GCRIT ) }$.
 Here
 one can not expand the correlation function $\corrf$ in
 the Taylor series in $G$ near $\GCRIT$.
 As in statistical mechanics, this example demonstrates
 the non-analytical dependence of correlation function on
 coupling constant.
 The $\GCRIT$ is the breaking point of $\corrf(G)$
 \begin{equation*}
  \corrf(G) =
  \begin{cases}
   0, & \mathrm{if}\  G \leqslant \GCRIT ; \\
   G \degen_1 \sqrt{1 - (\GCRIT/G)^2}, &
         \mathrm{if}\  G > \GCRIT .
  \end{cases}
 \end{equation*}
 For very large $G$, $G \gg (E_2 - E_1)/(2 \degen_1 -1 )$,
 \begin{equation*}
 \corrf = G \degen_1
 \sqrt{ 1 - \frac{(E_2 - E_1)^2}{ (2 \degen_1 -1 )^2 \, G^2 }}
 \approx
  G \degen_1 \left ( 1 - \frac{1}{2}
  \frac{(E_2 - E_1)^2}{ (2 \degen_1 -1 )^2 \, G^2 } + \ldots \right ).
 \end{equation*}
 The first term of the expansion in powers of $\GCRIT/G$ does not depend
 on the difference  $(E_2 - E_1)$ and coincides with the correlation
 function of one subshell system having
  $\degen_0 = 2 \degen_1$ and $p_0 = 1/2$.

 The calculated critical value of the interaction constant
 turned out to be proportional to the ratio $1/(2 \degen_1 - 1)$.
 It is not clear whether $\gcrit$ depends on the particle
 number or on the number of vacancies in the system.

 \remar
 Both normal
   ($w_1 = 1$, $w_2 = 0$, $t_2 = t_1 = 0$)
 and superconducting
 ($w_1 = 1/2 + \GCRIT/(2 G)$, $w_2 =  1/2 - \GCRIT/(2 G)$, $t_2 = t_1 > 0$)
 solutions are possible for  $G > \GCRIT$ in the considered example.
 The difference of their energies is easily calculated by
 Eq.\,(\ref{eq:averagedens}) with $\chemp = 0$,
 \begin{equation*}
 \langle \vert H \vert \rangle _{w_2 = 0}
 - \langle \vert H \vert \rangle_{w_2 > 0}
 =
  \degen_1 \left ( \degen_1 - \frac{1}{2} \right )
  G \left ( 1 - \frac{\GCRIT}{G} \right )^2 .
 \end{equation*}
 The subscripts $w_2 = 0$ and  $w_2 > 0$ indicate that the Hamiltonian
 is averaged over the wave function of either the normal state
 or the superconducting one, respectively.
 One can see that for $G > \GCRIT$ the energy of the normal state
 exceeds the energy of the superconducting one.

 \subsubsection{$\degen_1 \neq \degen_2$ and $\npart = 2 \degen_1$}%
 \label{subsect}
 Let us figure out how the value of the critical constant
 depends on the number of particles in the closed subshell $E_1$
 and on the number of empty single particle states in the
 subshell $E_2$.

 For $\npart = 2 \degen_1$ equation (\ref{eq:twosubparticles}) is
 \begin{equation}
 \label{eq:filledone}
  2 ( \degen_1 \xi_1 + \degen_2 \xi_2 ) = \degen_1 - \degen_2
 \quad \mathrm{or} \quad
  \degen_1 ( \xi_1 - \frac{1}{2} ) +
  \degen_2 ( \xi_2 + \frac{1}{2} ) = 0 .
 \end{equation}
 To simplify the calculations, we introduce
 the new variables $\delta_1$ and $\delta_2$ such as
 \begin{equation*}
  \xi_1 = \frac{1}{2} - \delta_1
 \quad \mathrm{and} \quad
  \xi_2 = - \frac{1}{2} + \delta_2 .
 \end{equation*}
 It follows from Eq.\,(\ref{eq:filledone}) that the unknown
 $\delta_1$ and $\delta_2$ are connected by the equation
 \begin{equation*}
  \degen_1 \delta_1 = \degen_2 \delta_2,
 \end{equation*}
 which can be taken into account by substitution
 \begin{equation*}
   \delta_1 = \degen_2 \delta ,
   \quad
   \delta_2 = \degen_1 \delta .
 \end{equation*}
 As $ -1/2 \leqslant \xi_{1,2} \leqslant 1/2$,
 the $\delta$ should be inside the interval
 \begin{equation*}
  0 \leqslant \delta \leqslant %
   \min \left( \frac{1}{\degen_1},  \frac{1}{\degen_2} \right) \,.
 \end{equation*}
 It follows from Eq.\,(\ref{eq:circletwo}) that
 \begin{equation*}
  \begin{aligned}
  t_1 & = \sqrt{\delta_1 (1 - \delta_1) } =
  \sqrt{\degen_2 \delta (1 - \degen_2 \delta)} \,,
 \quad \xi_1 = \frac{1}{2} - \degen_2 \delta \,, \\
  t_2 & = \sqrt{\delta_2 (1 - \delta_2) } =
  \sqrt{\degen_1 \delta (1 - \degen_1 \delta)} \,,
  \quad \xi_2 = - \frac{1}{2} + \degen_1 \delta \,. \\
  \end{aligned}
 \end{equation*}

 The unknown $\xi_{1,2}$ and $t_{1,2}$ expressed
 in terms
 of $\delta$ can be substituted into Eq.\,(\ref{eq:dimtwo}),
 the chemical potential can be excluded and the algebraic
 equation of sixth degrees can be obtained.
 The analysis of solutions of these equations is prohibitively
 difficult, and we narrow the problem by looking for values
 of the interaction constant $G$ at which the superconducting
 correlations will start to form.
 In other words, the unknown $\delta$ will be an infinitely
 small positive number.

 The abnormal densities
 $t_{1,2} \approx \sqrt{\degen_{2,1} \delta}$
 for small $\delta$.
 Therefore, the first of Eqs.\,(\ref{eq:dimtwo}) changes into
 \begin{equation*}
  - \rchem \sqrt{\degen_2 \delta}
  + g (\degen_1 - 1) \sqrt{\degen_2 \delta} \xi_1
  + g \degen_2 \sqrt{\degen_1 \delta } \xi_1 = 0 .
 \end{equation*}
 Since we consider $\delta > 0$ ($\delta = 0$
 corresponds to the normal solution),
 we can divide both sides of the equation
 by $\sqrt{ \degen_2 \delta}$.
 We transform the second equation of Eqs.\,(\ref{eq:dimtwo})
 in a similar way and obtain the simplified equations
 \begin{equation*}
  \begin{aligned}
  - \rchem
  + g \left ( \degen_1 - 1
  + \sqrt{\degen_1 \degen_2 } \right ) \xi_1 & = 0 , \\
  1 - \rchem + g \left ( \sqrt{\degen_1 \degen_2 }
   +  \degen_2 - 1 \right ) \xi_2  & = 0 .
   \end{aligned}
  \end{equation*}
 The linear equation for $\delta$ follows:
 \begin{multline}
 \label{eq:twodiflevdlt}
 \left \lbrack \left ( \degen_1 - 1 + \sqrt{\degen_1 \degen_2}
 \right ) \degen_2 +
  \left ( \degen_2 - 1 + \sqrt{\degen_1 \degen_2}
 \right ) \degen_1 \right \rbrack \delta \\
  \hbox{}\qquad =
  \frac{1}{2}
  \left ( \sqrt{\degen_1} + \sqrt{\degen_2} \right )^2 - 1 - \frac{1}{g} \, .
 \end{multline}
 The coefficient for $\delta$ in the left-hand side of the equation
 is independent of $g$ and is positive because
 $\degen_{1,2} \geqslant 1$.
 Therefore,  $\delta$ is positive if the right-hand side of the
 equation is positive.
 We rewrite the right hand side as
 \begin{equation*}
  \frac{1}{2}
  \left ( \sqrt{\degen_1} + \sqrt{\degen_2} \right )^2 - 1 - \frac{1}{g}
  = \frac{1}{\gcrit(\degen_1, \degen_2)} - \frac{1}{g}
  = \frac{g - \gcrit (\degen_1,\degen_2)}{\gcrit (\degen_1,\degen_2) \, g} \, ,
 \end{equation*}
 here
 \begin{equation}
 \label{eq:twolevelcrit}
  \displaystyle \gcrit(\degen_1, \degen_2) =
    \frac{1}{ \frac{1}{2} \left ( \sqrt{\degen_1} + \sqrt{\degen_2}
   \right )^2 - 1 } .
 \end{equation}
 Thus, $\delta$ will be an infinitesimal positive number if
 \begin{equation*}
   0 < g - \gcrit(\degen_1, \degen_2) \ll
  g \gcrit(\degen_1, \degen_2) .
 \end{equation*}
 The expression for $\gcrit(\degen_1, \degen_2)$ shows
 that the particle number (the size of the completely filled
 subshell $2 \degen_1$)
 and the vacancy number of the completely empty subshell
 ($2 \degen_2$) equally affect the critical value
 of the interaction constant.
 Such a somehow unexpected result can be explained by the
 observation that the Hamiltonian (\ref{eq:hamil}) has
 equal matrix elements for the processes of particle pair
 creation and destruction inside the $E_1$ subshell and
 the pair creation in the $E_2$ subshell.

 Now we calculate
 the correlation energy, chemical potential and particle
 number variance for infinitely small $\delta$, in other
 words, for $G$ satisfying the inequalities:
 \begin{equation*}
   0 < G - \GCRIT(\degen_1, \degen_2)
  \ll \gcrit(\degen_1, \degen_2) G ,
   \quad \mathrm{where}
   \quad \GCRIT(\degen_1, \degen_2) = \left ( E_2 - E_1 \right )
   \gcrit(\degen_1, \degen_2) .
 \end{equation*}
 The $\delta$ is calculated from Eq.\,(\ref{eq:twodiflevdlt}),
 afterwards $\delta_1$ and $\delta_2$ are determined
 \begin{equation*}
  \begin{aligned}
  \delta_1 & =  \frac{\degen_2}{R(\degen_1,\degen_2)}
   \left ( \frac{1}{\gcrit(\degen_1, \degen_2)}
  - \frac{1}{g} \right )
   =  \frac{\degen_2}{R(\degen_1,\degen_2)}
    \frac{g - \gcrit(\degen_1, \degen_2)}
     {g \, \gcrit(\degen_1, \degen_2) } , \\
  \delta_2 & =  \frac{\degen_1}{R(\degen_1,\degen_2)}
  \left ( \frac{1}{\gcrit(\degen_1, \degen_2)}
  - \frac{1}{g} \right )
   =  \frac{\degen_1}{R(\degen_1,\degen_2)}
    \frac{g - \gcrit(\degen_1, \degen_2)}
    {g \, \gcrit(\degen_1, \degen_2) } , \\
  {} & \mathrm{where} \quad
   R(\degen_1,\degen_2 ) =  \left ( \degen_1 - 1 +
   \sqrt{\degen_1 \degen_2} \right ) \degen_2 +
   \left ( \degen_2 - 1 + \sqrt{\degen_1 \degen_2}
  \right ) \degen_1 \,.
  \end{aligned}
 \end{equation*}
 The correlation energy is
 \begin{equation*}
  \begin{aligned}
  \corrf & = G \left ( \degen_1 t_1 + \degen_2 t_2 \right )
    \approx G  \left ( \degen_1 \sqrt{\delta_1}
    + \degen_2 \sqrt{\delta_2} \right ) = \\
  & = G \left ( \sqrt{\degen_1} + \sqrt{\degen_2} \right )
  \sqrt{ \frac{ \degen_1 \degen_2 \left ( g - \gcrit(\degen_1,\degen_2) \right )}%
  {R (\degen_1, \degen_2 ) \, g \, \gcrit (\degen_1,\degen_2) } } .
  \end{aligned}
 \end{equation*}
 Evidently, $\corrf$ can not be expanded in a Taylor
 series around $\gcrit$.
 The chemical potential is
  \begin{multline*}
 \chemp = E_1 + \frac{1}{2} \left (\degen_1 +
 \sqrt{\degen_1 \degen_2} - 2 \right ) G \\
 + \frac{\left ( \degen_1 + \sqrt{\degen_1 \degen_2} - 1 \right )
   \degen_2}{R (\degen_1, \degen_2 ) }
  \left ( 1 - \frac{g}{\gcrit(\degen_1,\degen_2)} \right ) (E_2 - E_1).
 \end{multline*}
 The particle number variance is
 \begin{equation*}
 \dispers =
 4 \left ( \degen_1 t_1^2 + \degen_2 t_2^2 \right ) =
   8
   \frac{\degen_1 \degen_2}{R(\degen_1, \degen_2 ) }
   \left ( \frac{1}{\gcrit} - \frac{1}{g} \right ) ,
 \end{equation*}
 therefore, $\dispers$ is proportional to the difference
 $\left ( g - \gcrit \right )$, if $g$ slightly exceeds $\gcrit$.

 \subsubsection{ $\npart < 2 \degen_1 $ }
 Let the number of particles be
 $\npart = 2 \degen_{1} - 2 P_1$,
 that is, the $P_1$ pair of particles is removed from the
 low closed subshell.
 We have shown in subsection \ref{sec:single} that the abnormal density
 $t_1$ is positive for any $G$ in this case.
 One needs to find out at what $G$ the abnormal
 density $t_2$ will be non-zero.
 Two variants are possible.
 Either $t_2 > 0$ at any positive coupling constant or
 a certain critical value $\tilde{G}$ exists such as
  $t_1 > 0$ and $t_2 = 0$ if  $0 < G < \tilde{G}$.

 With $\npart = 2 \degen_{1} - 2 P_1$
 particle number equation (\ref{eq:twosubparticles}) is
 \begin{equation*}
  2 \degen_1 \xi_1 + 2 \degen_2 \xi_2  = \degen_1 - 2 P_1 - \degen_2
  \quad \mathrm{or} \quad
     \degen_1 \left ( \xi_1 - \frac{1}{2} + p_1 \right ) + \degen_2
    \left ( \xi_2 + \frac{1}{2} \right ) = 0.
 \end{equation*}
 Here $p_1 = P_1/\degen_1$.
 We put
 $\xi_1 = 1/2 - p_1 - \delta_1$ and $\xi_2 = - 1/2 + \delta_2$
 and obtain
 \begin{equation*}
  t_1 = \sqrt{ \left ( p_1 + \delta_1 \right ) \left (
       1 - p_1 - \delta_1 \right ) }
  \quad \mathrm{and} \quad
  t_2 = \sqrt{  \delta_2  \left ( 1 - \delta_2 \right ) } .
 \end{equation*}
 New variables $\delta_1$ and $\delta_2$
 are related to each other by
 $ \degen_1 \delta_1 = \degen_2 \delta_2 $,
 which can be satisfied if
 $\delta_1 = \degen_2 \delta$ and $\delta_2 = \degen_1 \delta$.
 The density $t_2 > 0$ if $\delta > 0$.
 The boundaries (\ref{eq:circleone}) of $\xi_{1,2}$
 lead to the inequalities
 \begin{equation*}
  0 \leqslant \delta \leqslant \min \left (
  \frac{1-p_1}{\degen_2}, \frac{1}{\degen_1} \right ) .
 \end{equation*}

 As in the previous example, Eqs.\,(\ref{eq:dimtwo}) can be
 transformed into an algebraic equation of the sixth degree
 with respect to $\delta$.
 We obtain simplified equations for infinitesimal positive
 $\delta$.
 For small $\delta$
 \begin{equation*}
   \begin{aligned}
  t_1 & = \sqrt{ p_1 ( 1 - p_1 )}
  + \frac{1}{2} \frac{1 - 2 p_1}{\sqrt{ p_1 ( 1 - p_1 )}}
    \degen_2 \delta + o(\delta) , \\
 t_2 & = \sqrt{\degen_1 \delta}
  \left ( 1 - \frac{1}{2} \degen_1 \delta + o(\delta) \right) . \\
   \end{aligned}
 \end{equation*}
 The symbol $o(x)$ stands for the functions of $x$ that
 $ \displaystyle \frac{o(x)}{x} \to 0 $ when $ x \to 0$.

 For $ 0 < \delta < 1 $ the inequalities
 $ 0 < \delta < \sqrt{\delta} < 1 $ are satisfied;
 therefore, the expansion of $t_{1,2}$ should be carried for
 $\sqrt{\delta}$ not for $\delta$ itself.
 Keeping the terms with zero and first powers of $\sqrt{\delta}$,
 we obtain the approximate
 \begin{equation*}
   \begin{aligned}
   t_1 & = \sqrt{  p_1   ( 1 - p_1 ) }  , \quad &
    \xi_1 & = 1/2 - p_1  , \\
   t_2 & = \sqrt{\degen_1 \delta } , &
     \xi_2 & = - 1/2 .
   \end{aligned}
 \end{equation*}
 The approximate equations follow from the exact
 ones (\ref{eq:dimtwo})
 \begin{gather*}
    -\rchem
        - g \left ( \frac{1}{2} - p_1 \right )
        + g \left( \degen_1
          + \degen_2
   \sqrt{\frac{\degen_1 \delta}{p_1 (1 - p_1)}}\, \right )
           \left ( \frac{1}{2} - p_1 \right )  = 0 , \\
   \left ( 1 - \rchem \right ) \sqrt{\degen_1 \delta}
    + \frac{g}{2} \sqrt{\degen_1 \delta}
    - \frac{g}{2} \left ( \degen_1  \sqrt{p_1 (1 - p_1)}
    + \degen_2 \sqrt{\degen_1 \delta} \right )  = 0 .
 \end{gather*}
 The solutions of the system of approximate equations are
 \begin{equation*}
  \begin{aligned}
    \rchem & = g \left ( \frac{1}{2} - p_1 \right )
       \left ( \degen_1 - 1 + \degen_2
       \sqrt{\frac{\degen_1 \delta}{p_1 (1 - p_1)}}\, \right ) , \\
   t_2 & = \sqrt{\degen_1 \delta}  =
 \frac{g}{2} \,
 \frac{ \degen_1 \sqrt{p_1 (1 - p_1) }}%
 {1 + g \left ( 1 + p_1 (\degen_1 - 1)
   - \left (\degen_1 + \degen_2\right )/2 \right )} . \\
  \end{aligned}
 \end{equation*}
 When calculating $t_2$,  we ignored the term
 proportional to $\degen_1\delta$ produced in the
 second equation from the product
  $\rchem\sqrt{\degen_1\delta}$.

 The obtained solutions show that the abnormal density $t_2$
 is the infinitely small positive number for
 any positive infinitely small $g$.
 The correlation function is approximately equal to
 \begin{equation*}
  \corrf \approx G \degen_1 \sqrt{p_1 (1 - p_1)}
  \left ( 1 +  \frac{G}{2 \left (E_2-E_1 \right)} \degen_2 \right ) .
 \end{equation*}
 The linear on $g$ part of $t_2$ is taken into account here.
 The chemical potential calculated with the same accuracy is
 \begin{equation*}
  \chemp \approx E_1 + G \left \lbrack
     \left (\frac{1}{2} - p_1 \right ) \degen_1
     + p_1 - 1 + \frac{G}{2 \left (E_2-E_1 \right )}
      \left (\frac{1}{2} - p_1 \right ) \degen_1 \degen_2
      \right \rbrack .
 \end{equation*}
 It is easy to see that the chemical potential calculated
 to the first order in $G$ is equal to the chemical potential
 for a system consisting of a single open subshell.

 \subsubsection{ $\npart = 2 \degen_1 + 2Q, \ 0 < Q < \degen_2 $ }
 Let the number of particles be sufficient to fill the subshell
 $E_1$ completely and the subshell $E_2$ -- partially.
 Equation (\ref{eq:twosubparticles}) gives
 \begin{gather*}
   2 \left ( \degen_1 \xi_1 + \degen_2 \xi_2 \right )
   = \degen_1 - \degen_2 + 2 Q , \\
  \degen_1 \left ( \xi_1 - \frac{1}{2} \right )
  + \degen_2 \left ( \xi_2 + \frac{1}{2} - q \right ) = 0,
  \quad q = \frac{Q}{\degen_2} .
 \end{gather*}
 We introduce $\varepsilon_{1,2}$ such that
 $\xi_1 = 1/2 - \varepsilon_1$ and  $\xi_2 = -1/2 + q + \varepsilon_2$.
 The new variables $\varepsilon_1$ and $\varepsilon_2$ are connected
 by the equation
 $\degen_1 \varepsilon_1 = \degen_2 \varepsilon_2$,
 which can be solved by substitutions
 $ \varepsilon_1 = \degen_2 \varepsilon $ and
 $\varepsilon_2 = \degen_1 \varepsilon $.

 As in the previous cases,
 in order to study the appearance of superconducting pair
 correlations, we keep the terms
 proportional to the zero and first degrees of infinitesimal
 $ \sqrt{\varepsilon} $ and obtain
 \begin{equation*}
  \begin{aligned}
   \xi_1 & = \frac{1}{2} - \degen_2 \varepsilon
    \approx \frac{1}{2} , & \quad
   t_1 & = \sqrt{ \degen_2 \varepsilon
      \left ( 1 -  \degen_2 \varepsilon \right )}
    \approx \sqrt{ \degen_2 \varepsilon } , \\
   \xi_2 & = - \frac{1}{2} + q + \degen_1 \varepsilon
    \approx - \frac{1}{2} + q , & \quad
  t_2 & = \sqrt{\left ( q + \degen_1 \varepsilon \right )
    \left ( 1- q - \degen_1 \varepsilon \right ) }
   \approx \sqrt{ q \left ( 1 - q \right )} .
  \end{aligned}
 \end{equation*}
 We substitute
 these approximate values into Eqs.\,(\ref{eq:dimtwo}),
 divide both sides of the second equation by non-zero $ t_2 $,
 and determine $\rchem$ and $t_1$ from the coupled
 linear equations
 \begin{gather*}
  \rchem \approx 1 + g \left ( q - \frac{1}{2} \right )
  \left \lbrack  \degen_2 - 1 + \degen_1 \frac{t_1}{t_2}
 \right \rbrack , \\
  t_1 \approx \sqrt{\degen_2 \varepsilon} = \frac{g}{2} \,
  \frac{ \degen_2 \sqrt{ q \left ( 1 - q \right )}}%
  { 1 + g \left \lbrack 1 + \left ( \degen_2 - 1 \right ) q
   - \left ( \degen_1 + \degen_2 \right ) / 2 \right \rbrack }
 \approx \frac{g}{2} \degen_2 \sqrt{q(1-q)} .
 \end{gather*}
 The obtained $t_1$ and $t_2$ are used for calculation of the
 correlation function and chemical potential
 \begin{equation*}
 \begin{aligned}
  \corrf & \approx G \degen_2 \sqrt{q \left ( 1 - q \right )}
  \left ( 1 + \frac{G}{2 (E_2 - E_1 )} \degen_1 \right ) , \\
 \noalign{\smallskip}
  \chemp & \approx E_2 + G \left \lbrack \left ( q - \frac{1}{2} \right ) \degen_2
  - q + \frac{G}{2 (E_2 - E_1 )} \left ( q - \frac{1}{2} \right )
  \degen_1 \degen_2 \right \rbrack .
  \end{aligned}
 \end{equation*}
 Please note that the correlation function and chemical potential
 for small $g$ reproduce the exact solutions for the single open
 subshell.

 The last two examples dealing with two-subshell systems
 having one open subshell show that both abnormal densities
 will be non-zero (the superconducting solutions exist for
 both subshells) at any small positive coupling constant.
 The chemical potential is found to be near the energy of the
 single particle states forming the open subshell.

 \section{Several subshells}
 Let the system have $\msubs$ subshells, $\msubs > 2$.
 We number the subshells so that $E_k \leqslant E_l$
 if $k < l$.
 In the model of independent particles, as the number
 of particles grows the subshells are gradually filled:
 from the subshells with lower single-particle energy to the
 subshells with larger ones.
 We denote by $F$ the number of the largest energy subshell
 which still has particles in it.
 Therefore, the total number of particles in the system is
 \begin{equation*}
   \npart = \sum_{k=1}^{F} 2 \degen_{k} - 2 P,
 \end{equation*}
 here $P$ is the number of particle pairs required to
 fill the subshell $F$ completely.
 If $P = 0$, we have the closed subshell system.
 If $ 0 < P < \degen_F $, the system is the open subshell system.

 The average number of particles (\ref{eq:averagepart})
 can be written as
 \begin{equation*}
  \sum_{k=1}^{F-1} \degen_k \left ( \xi_k - \frac{1}{2} \right )
 + \degen_F \left ( \xi_F - \frac{1}{2} + p_F \right )
 + \sum_{k=F+1}^{\msubs} \degen_k \left ( \xi_k + \frac{1}{2} \right )
 = 0.
 \end{equation*}
 Here $p_F = P /\degen_F$.
 Instead of variables $\xi_k$ we introduce new unknowns $\delta_k$
 \begin{equation*}
  \xi_k =
  \begin{cases}
  \frac{1}{2} - \delta_k, & 1 \leqslant k < F; \\
  \frac{1}{2} - p_F - \delta_k, & k = F ; \\
  - \frac{1}{2} + \delta_k, & F < k \leqslant \msubs .
  \end{cases}
 \end{equation*}
 They are connected by
 \begin{equation}
 \label{eq:partmanylevels}
  \sum_{k=1}^{F} \degen_k \delta_k
 =   \sum_{l=F+1}^{\msubs} \degen_l \delta_l.
 \end{equation}
 The conditions $-1/2 < \xi_k < 1/2$, valid for any $k$,
 dictate the inequalities for $\delta_k$:
 \begin{equation*}
  0 < \delta_k < 1 \ \mathrm{for} \ k \neq F,
 \ \mathrm{and} \ -p_F < \delta_F < 1 - p_F .
 \end{equation*}
 Please note that now $\delta_F$ can be not only a positive
 but also a negative number.
 It follows from Eq.\,(\ref{eq:circletwo}) that
 the abnormal densities are
 \begin{equation*}
  t_k =
  \begin{cases}
   \sqrt{\delta_k  ( 1 - \delta_k )}, &  k \neq F ; \\
 \noalign{\smallskip}
   \sqrt { ( p_F + \delta_F ) ( 1 - p_F - \delta_F  )},
   & k = F .
  \end{cases}
 \end{equation*}
 Equations (\ref{eq:main}) can be written as the system
 \begin{equation}
 \label{eq:manylevels}
  \begin{aligned}
  \left \lbrack E_k - G \left ( 1 - \delta_k \right )
   - \chemp \right \rbrack t_k
  + G D \left ( \frac{1}{2} - \delta_k \right ) &
  = 0 \,, \quad & 1 \leqslant k < F , \quad \\
  \left \lbrack E_F - G \left ( 1 - p_F - \delta_F \right )
  - \chemp \right \rbrack t_F
  + G D \left ( \frac{1}{2} - p_F - \delta_F \right ) &
  = 0 \,, \quad & {} \\
  \left \lbrack E_l - G  \delta_l - \chemp \right \rbrack t_l
  - G D \left ( \frac{1}{2} - \delta_l \right ) &
  = 0 \,, \quad & F < l \leqslant \msubs \,.
  \end{aligned}
 \end{equation}
 that connects
 $\chemp$, $\delta_{k}$ and $t_k$,
 $k = 1, \ldots , \msubs$.

 \subsection{$ p_F > 0 $}
 For the open subshell nuclei $p_F > 0$.
 We are looking for the conditions under which all
 $\delta_k$ with $k \neq F$ will be positive
 infinitely small numbers.
 If follows from Eq.\,(\ref{eq:partmanylevels}) that
 $\delta_F$ will also be an infinitesimal number of the same
 order as $\delta_k$.
 The abnormal densities are
 \begin{equation*}
   \begin{aligned}
  t_F & = \sqrt{ p_F ( 1 - p_F )}
  + \frac{1}{2} \frac{1 - 2 p_F}{\sqrt{ p_F ( 1 - p_F )}}
    \delta_F + o(\delta_F) , \\
 t_k & = \sqrt{\delta_k}
  \left ( 1 - \frac{1}{2} \delta_k + o(\delta_k) \right),
  \quad k \neq F . \\
   \end{aligned}
 \end{equation*}
 For $\delta_k$ with $k \neq F$ the inequalities
 $ 0  < \delta_k < \sqrt{\delta_k} $ are satisfied;
 therefore, we consider $\sqrt{\delta_k} $ as having
 the first order of smallness and will keep
 $ \sqrt{\delta_k} $ in the zeroth and first degrees.
 As a result, we have
 \begin{gather*}
   t_k \approx \sqrt{\delta_k}, \quad k \neq F , \\
   t_F \approx \sqrt{p_F \left ( 1 - p_F \right )} , \\
   D \approx \degen_F \sqrt{p_F \left ( 1 - p_F \right )}
      + \sum_{ l \neq F } \degen_l \sqrt{\delta_l}
    \approx \degen_F \sqrt{p_F \left ( 1 - p_F \right )} .
 \end{gather*}
 The terms proportional to $\degen_l \sqrt{\delta_l}$ are
 neglected in the last expression because they are infinitely
 small in comparison with the finite term
 $\degen_F \sqrt{p_F \left ( 1 - p_F \right )}$.
 We substitute the approximate expressions into Eq.\,(\ref{eq:manylevels})
 and obtain
 \begin{gather*}
  \chemp = E_F + \left \lbrack \left (\frac{1}{2} - p_F \right )
  \degen_F  - ( 1 - p_F ) \right \rbrack G, \\
 \noalign{\smallskip}
  t_k = \frac{1}{2} \degen_F \sqrt{p_F (1 - p_F)} \frac{G}{E_F-E_k}
   + o \left(\frac{G}{E_F-E_k} \right ) ,
  \quad 1 \leqslant k < F,  \\
 \noalign{\smallskip}
  t_l = \frac{1}{2} \degen_F \sqrt{p_F (1 - p_F)} \frac{G}{E_l-E_F}
    + o \left(\frac{G}{E_l-E_F} \right ) ,
  \quad F < l \leqslant \msubs .
 \end{gather*}
 The correlation function and the particle number variance
 are
 \begin{gather*}
 \corrf \approx G \, \degen_{F} \sqrt{p_F (1 - p_F)} \left (
 1 + \frac{1}{2}\sum_{k = 1}^{F - 1} \degen_k \frac{G}{E_F-E_k}
   + \frac{1}{2}\sum_{ l = F + 1}^{\msubs}
     \degen_l \frac{G}{E_l-E_F} \right ) , \\
 \noalign{\smallskip}
 \dispers \approx \degen_{F} \, p_F (1 - p_F) \left \lbrack 4
  + \sum_{k=1}^{F-1} \degen_F  \degen_k
  \left( \frac{G}{E_F-E_k} \right )^2
  + \sum_{l=F+1}^{\msubs} \degen_F  \degen_l
  \left( \frac{G}{E_l-E_F} \right )^2
  \right \rbrack .
 \end{gather*}
 For a small coupling constant both the correlation function
 and the particle number variance are mostly determined by the
 abnormal density of the open subshell.
 The following inequalities are valid:
 \begin{equation*}
 \corrf > G \, \degen_{F} \sqrt{p_F (1 - p_F)}
 \quad \mathrm{and} \quad
 \dispers > 4 \degen_{F} \, p_F (1 - p_F) .
 \end{equation*}
 We would like to note that the particle number variance
 $ \dispers $ is bounded below by the positive number
 when $G \to 0$.

 One can see that the interaction spreads the influence of
 non-zero $t_F$ over all subshells in the system.
 The abnormal densities differ from zero at all subshells
 enveloped by the interaction.
 In the system of like nucleons with the open subshell the
 superconducting solution exists at any
 arbitrarily small constant of attractive interaction.

 \subsection{$ p_F = 0 $}
 Now we consider the closed subshell system with $p_F = 0$.
 Let us assume for a moment
 that all $E_k$ with $1 \leqslant k < F$
 are equal to $E_F$ and
 all $E_l$ with $F < l \leqslant \msubs $ are equal to $E_{F+1}$.
 By this assumption we return to the two-subshell problem with
 \begin{equation*}
  \widetilde{\degen}_{F} = \sum_{k = 1}^{F} \degen_k
  \quad \mathrm{and} \quad
  \widetilde{\degen}_{F+1} = \sum_{l = F+1}^{\msubs} \degen_l .
 \end{equation*}
 We have shown in subsection \ref{subsect} that in the present case
 the superconducting pair correlations start to appear
 if the coupling constant $G$ exceeds the threshold value
 (\ref{eq:twolevelcrit}) which is here
 \begin{equation*}
   \GCRITp =
   \frac{ E_{F+1} - E_F }{\frac{1}{2}
    \left ( \sqrt{\widetilde{\degen}_{F}}
   + \sqrt{\widetilde{\degen}_{F+1}} \right )^2 - 1} .
 \end{equation*}
 Please note that $2 \widetilde{\degen}_{F} = \npart$.
 During the development of the superconducting correlations,
 pairs of particles begin to jump from fully occupied
 subshells into free subshells.
 The difference $ \left (E_{F+1} - E_F \right ) $ is the
 lowest energy of such transitions.
 Therefore, in the initial $\msubs$-subshell system
 the constant $\GCRITp$ gives the lower bound for the
 actual critical value of the interaction constant $\GCRIT$.

 On the other hand,
 \begin{equation*}
   \GCRITpp =
   \frac{ E_{F+1} - E_F }{\frac{1}{2}
  \left ( \sqrt{\degen_{F}}
   + \sqrt{\degen_{F+1}} \right )^2 - 1} ,
 \end{equation*}
 is the critical constant for the system with only two
 interacting subshells taken into account, and the contributions
 of other subshells are ignored.
 Therefore, $\GCRITpp$ gives the upper bound for $\GCRIT$.

 This consideration shows that in the closed subshell system
 only
 the normal solutions are possible if the interaction constant
 satisfies $0 < G < \GCRITp$.
 All abnormal densities are equal to zero in this case.
 If $G$ exceeds $ \GCRITpp $, the superconducting
 solution is possible and all abnormal densities become positive
 numbers.
 The intermediate case with $\GCRITp < G < \GCRITpp $
 requires additional study.

 \remar %
 We have used the simplest model Hamiltonian (\ref{eq:hamil})
 with the constant attractive interaction.
 The realistic Hamiltonian can be written \cite{VBZ2001} as
 \begin{multline*}
  H
  = \sum_k \sum_{m_k = -j_k}^{j_k} E_k a^{\dag}_{k, m_k} a_{k,m_k} - \\
  - \frac{1}{4} \sum_{k,l} G_{k,l}
 \sum_{m_k=-j_k}^{j_k} \sum_{m_l=-j_l}^{j_l}
 (-1)^{j_k - m_k}  a^{\dag}_{k, m_k} a^{\dag}_{k, -m_k}
 \, (-1)^{j_l - m_l}  a_{l, -m_l} a_{l, m_l} .
 \end{multline*}
 If all matrix elements $G_{k,l}$ are positive numbers,
 our qualitative conclusions about the system with open subshell
 will survive.
 The expressions for $\chemp $ and $\corrf$ will be
 more complicated of course.

 \remar
 We have considered the spherical nuclei.
 To obtain formulae for deformed nuclei, one should
 put $j_k = 1/2$ and $\degen_k = 1$ for all subshells.
 The formulae obtained for the closed subshell spherical nuclei
 are suitable for deformed nuclei.
 The deformed nucleus can be open subshell nucleus only
 if the energies $E_F$ and $E_{F+1}$ coincide.

 \section*{Conclusion}

 We have considered the appearance of superconducting pair
 correlations in spherical even-even nucleus using the
 simplest model Hamiltonian.
 The influence of the monopole pairing interaction on
 the energy of single-particle states was taken into account.

 It is shown that the emergence of pair correlations
 depends on the particle number and shell structure.

 In the open subshell system non-zero abnormal densities appear
 for any small positive coupling constant.
 The new result obtained in the present paper is that
 at infinitely small positive $G$ the
 abnormal densities differ from zero at each subshell
 participating in the pairing interaction.

 In the closed subshell system, superconducting pair correlations
 begin to form when the coupling constant exceeds a certain
 critical value $\GCRIT$.
 If the coupling constant is less than $\GCRIT$,
 the normal solution is the only solution.
 All abnormal densities and correlation function equal to zero.
 The rough lower and upper bounds of $\GCRIT$ are obtained.
 More accurate estimations of $\GCRIT$ are required.
 Both normal and superconducting solutions are possible
 for $G$ larger than $\GCRIT$.


\begin{thebibliography}{i}
 \bibitem{Bel1959} %
  {\it Belyaev S.\,T.\/} Effect of pairing correlations on
  nuclear properties //
  Mat.\ Fys.\ Medd.\ Dan.\ Vid.\ Selsk.\ 1959.\ 31.\ No.\ 11.
%
 \bibitem{VGS1971}
  {\it Soloviev V.\,G.\/}  Teoriya sloghnyh yader. // 1971. Nauka. GRFML. M.
  [English translation: Theory of Complex Nuclei. 1976. Pergamon Press.]
%
 \bibitem{BCS50} %
  Fifty Years of Nuclear BCS: Pairing in Finite Systems /
  Ed.: Broglia R.\,A. and Zelevinsky V.:
  World Scientific Publishing Co.\ Pte.\ Ltd.\ 2013.
%
 \bibitem{VBZ2001}
 {\it Volya A., Brown B.\,A., Zelevinsky V.\/}
 Exact solution of the nuclear pairing problem //
 Phys.\ Lett.\ B.\ 2001.\ V.\ 509.\ P.\ 37.

 \end{thebibliography}
\end{document}